.

# A Novel Composite Resilience Indicator for Decentralized Infrastructure Systems (CRI-DS)

Lamis Amer*, Murat Erkoc*[1], Esber Andiroglu+ & Nurcin Celik*

*Department of Industrial and Systems Engineering, University of Miami FL

+Department of Environmental and Architectural Engineering, University of Miami FL

## Abstract

Resilience is a key driver for planning adaptation strategies to mitigate risks due to both natural and anthropogenic hazards. The effectiveness of a resilience-driven decision-making strategy for adapting systems against stressors depends on how resilience is mapped to decision variables. This requires a functional resilience metric, without which it would not be possible to identify the priority needs for improvement, assess changes, or show improvement in post-adaptation resilience. In this paper, we aim to contribute to existing methodologies by proposing a novel model-based resilience assessment strategy while building on the notion of composite resilience indicators. Our proposed indicator is functional, reproducible, and is mapped to adaptation decisions. We propose a framework for integrating the developed resilience functional form into an adaptation decision-making model. To illustrate our approach, we use a case study on assessing resilience of On-Site Wastewater Treatment and Disposal Systems (OSTDS) exposed to risks due to sea-level rise.

## 1. Introduction and Background

Under an incessantly warming climate, global sea levels have been rising and are projected to continue to rise in the future (Blunden and Arndt 2018). According to the IPCC 6th Assessment Report[2], under the intermediate greenhouse gas emissions scenario, global sea levels are projected to rise by 0.56 m $\pm$ 0.2 by 2100. Rising sea levels constitute threats to living communities with increasing risk of coastal flooding events. The risks and challenges presented by SLR are not limited to inundation. Higher water levels also modify shoreline hydrodynamics, drive long-term coastal erosion, and increase saltwater intrusion into coastal freshwater aquifers. Every increment of SLR leads to greater nuisance flooding and impeded drainage, loss of lands, shifts in habitat types, contaminated freshwater supplies, and potential water quality impairment. These myriad challenges have prompted practitioners and researchers to develop models and sustainable solutions that can mitigate the adverse effects of rising seas. Planning the pathway to building communities that are both sustainable and resilient necessitates holistic and proactive approaches that not only focus on risk reduction but also provide solutions for long-run adaptation to changing environment. Thus far, the majority of solutions have been narrow in scope, focusing on particular problems or risks with ad hoc, disaster-driven, and reactive views. Recent studies give emphasis to a call for proactive, threat-driven, and mitigative focus in modeling and decision making for adaptation (Alyamani, Damgacioglu et al. 2016; Park 2022). This call has fueled the rising interest in research tackling urban resiliency needs.

---

[1] Corresponding author. Email: merkoc@miami.edu

[2] https://sealevel.nasa.gov/ipcc-ar6-sea-level-projection-tool?type=global

According to the city resilience framework presented by Arup & The Rockefeller Foundation (Silva and Morera 2014), a critical factor that contributes to community resilience is the effective provision of critical services, such as water/wastewater utilities, energy, communications, transportation etc. These services are provided through man-made infrastructure systems and utilities that function interdependently and become critical if their incapacity or disruption has a debilitating impact on the physical ability, economic stability, and safety of the society. Moreover, a failure in such a system, or the loss of its services, could cascade across boundaries causing failures in multiple infrastructures with potentially catastrophic consequences. The majority of critical infrastructure systems possess a network-based structure. The architecture of these networks shapes the system's ability to respond and adapt to possible disruptions. Therefore, many of the resilience metrics proposed in the literature rely on different indices adopted from graph theory, such as accessibility and connectivity to assess the extent of service disruption and impact propagation in the event of failure of a link or node, as proposed by Balijepalli and Oppong (2014), Demirel, Kompil et al. (2015), Li, Kou et al. (2020), Maiolo, Pantusa et al. (2018), Pinto, Varum et al. (2010) and Suarez, Anderson et al. (2005). These measures apply to all network-based infrastructure systems, including transportation, water supply, wastewater networks, power transmission, and telecommunication networks. However, distinct metrics need to be adopted when assessing the resilience of critical, decentralized infrastructure systems, such as agricultural fields, fisheries, and onsite wastewater treatment and disposal systems.

Different taxonomies are introduced to review and classify quantitative resiliency measures by recent studies such as Beccari (2016) and Hosseini, Barker et al. (2016). In general, the majority of the proposed measures can be grouped into two approaches: functional or performance-based resilience, and structural resilience, also known as system-driven, built-in, and engineering resilience (Henry and Ramirez-Marquez, 2012). Functional resilience or the resilience trapezoid method is proposed in the literature with initial application in seismic-related hazards (Bruneau, Chang et al., 2003). This method captures the time-dependent performance measures during the degradation and recovery phases of a system under threat, resulting in the multi-phase curvature of the resilience trapezoid. Provided sufficient historical damage data is available, the resilience trapezoid can be generated by simulation (Ghosh, Mohanta et al., 2021), or predicted based on probabilistic-damage and fragility curves, loss functions, and recovery curves of the system under investigation. In many cases, data may not be available to model and/or predict the shape of the resilience trapezoid. Furthermore, the development of resilience measures based on system outputs may provide limited insight as to what constitutes a resilient system as it fails to capture the dynamics within the considered system and between the system and its surrounding environment. Therefore, process measures and model-based assessments may be powerful in assessing the resilience of dynamic processes and systems (Woltjer 2008).

To bridge these gaps, particular attention has been given to the development of composite indices to measure vulnerability and resilience (Beccari 2016), such as in the disaster-focused, composite Social Vulnerability Index (SoVI) (Cutter, Boruff et al. 2003). The development of composite indicators, on one-hand, fulfills the need for process-driven resilience measures, where the inherent systems' characteristics, structure, and dependencies with the surroundings shape their resilience. On the other hand, it provides a methodology to operationalize resilience while hedging against data deficiency.

The process for developing a composite resilience indicator consists of three main steps, (1) identifying indicators, sub-indicators, and variables, (2) data manipulation and transformation, and (3) aggregation. In order to generate the overall resilience metric most of the proposed aggregation strategies rely on the application of fuzzified rules such as min-max IF-THEN rules for conjunctive (AND) and disjunctive (OR) reasoning (Kammouh, Noori et al. 2018, Schaefer, Thinh et al. 2020), weighted multi-criteria overlay analysis (Afrasiabi, Chalmardi et al. 2022), or Analytic Hierarchy Process (AHP) (Ataoui and Ermini 2015). The fuzzy inference models generally specify explicit conditions for the inputs to generate an output. These conditions hinder the overlay strategy from robustness, while also aggregating the same importance to all fuzzified factors. Addressing this issue, AHP and weighted multi-criteria overlay analysis are mainly utilized to prioritize criteria based on assigned weights. However, decision making based on weighted comparisons in most part rely on the judgments of the decision maker and often results in the assignment of arbitrary weights. Moreover, in the AHP, since the main goal is to order factors in their importance, pairwise comparisons must be established for determining the weights. This process becomes computationally expensive with a potentially large number of parameters $n$, where the number of paired comparisons that need to be repeated with every iteration of the decision-making model is $n^2$. The complexity of the process is further exacerbated by the selection of scale and range for the weights from an arbitrary spectrum. As a resolution, building on the theories of reliability engineering, we propose a reproducible and systematic parameter aggregation strategy for computing the overall resilience index for each system.

The contribution of this study to the literature on operationalizing resilience is threefold: (i) introduction of a Composite Resilience Indicator for Decentralized Systems (CRI-DS), that is reproducible and modeled as a function of the adaptation decision variables, (ii) assessment of the resilience of On-Site Wastewater Treatment and Disposal Systems (OSTDS) to risks due to sea-level rise, and (iii) reflection of the intra-dependencies between infrastructure systems (with freshwater and wastewater resources) into the proposed functional resilience metric. We also propose a framework for integrating the proposed resilience measure into adaptation decision-making by linking adaptation decisions and the system response capacities under an optimization model. Another novel and tangible deliverable of this research is the production of a sharable geospatial model that can be utilized to integrate and visualize the proposed CRI-DS. We employ a case study based on 107,000 active OSTDS in Miami-Dade County in Florida to demonstrate the functionality of our proposed resilience metrics approach and its integration into the geospatial dataset. The resilience indicators associated with each of these OSTDS are computed based on the proposed approach and mapped into the geospatial model to facilitate additional analysis of the dataset. The dataset includes attributes quantifying each of the system capacities shaping resilience, namely, the resistive, adaptive, and restorative capacities. Once integrated into the policymaking mechanisms, this dataset can provide valuable insights in guiding and prioritizing adaptation decisions.

The rest of the paper is organized as follows. In Section 2, we discuss our proposed resilience assessment model in relation to the OSTDS. This section is divided into four sub-sections, each representing a particular step in the proposed model. In sub-section 2.1, we discuss the exploratory research phase followed by the significance of systems thinking and causal relationships in understanding how various system characteristics may shape the likelihood of failure and recovery, hence its resilience. In sub-section 2.2, we detail our data collection and interpretation process, the role of spatial analysis in understanding how the OSTDS systems and identified factors interact with the surrounding environments and other critical infrastructure systems, and how these interactions shape OSTDS systems' ability to respond to risks. For instance, the proximity of OSTDS to existing wellheads shapes their criticality in the event of failure where there is a higher likelihood of impact propagation as compared to other far-located systems. Integrating these identified factors to constitute the resilience index formula is detailed in sub-sections 2.3 and 2.4, where we discuss our methodology to normalize and transform the individual measures and generate the overall resilience index, respectively. In Section 3 we present a framework for embedding the proposed resilience index into adaptation decision-making. Finally, Section 4 concludes the paper.

## 2. Proposed Resilience Assessment Model

Capturing resilience effectively necessitates a clear understanding of what factors make up a resilient system, and how these factors coalesce into the state and functioning of the system. In order to construct hypotheses on how the system behaves under current and future sea-levels, we start by conducting an exploratory research where we investigate all direct and indirect relationships between various system failure modes and risks evolving as a result of sea-level rise. Then a set of system-related attributes, which may be critical in driving the system's response and restorative capacities and thus its resilience, are identified.

Because these attributes might reflect various system characteristics and response capacities, they can be measured on different scales (i.e., distance, percentages, ratios, etc.). Moreover, the roles played by each attribute in shaping the overall system resilience are not equally important. Responding to this need, our model adopts multiple fuzzy transformation techniques to generate normalized weighted values between 0 and 1 depending on the significance of the parameter considered. Upon transformation, in order to generate the CRI-DS we propose an aggregation methodology that builds on the concepts of reliability engineering. This phase results in the proposed functional CRI-DS and a sharable geospatial dataset of existing septic systems associated with the estimated resilience index. Figure 1 below summarizes the resilience assessment model steps. The rational and mechanism for each of these steps are elaborated in the following subsections of this paper.

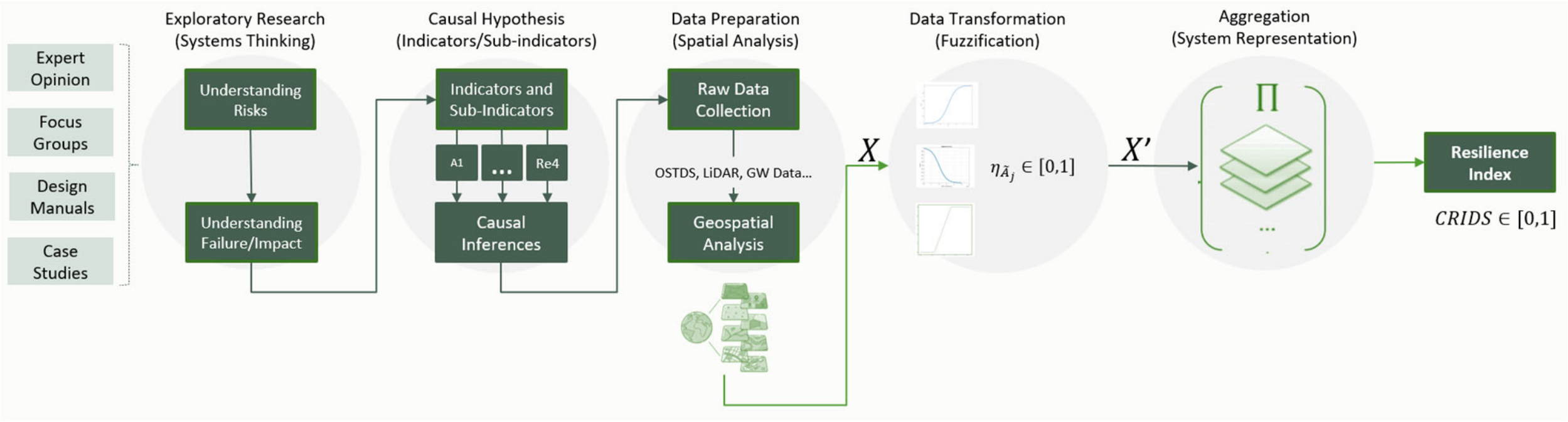

***Figure 1.*** *The Resilience Assessment Model*

### *2.1. Exploratory Research: Systems Thinking*

Systems thinking has its foundation in the field of systems dynamics, first introduced by Forrester (1958). It allows a comprehensive understanding of systems by focusing on how the different system components interact with each other and with the external environment in order to generate a certain behavior. This makes systems thinking an effective approach to model and solve problems related to complex and dynamic systems. For this, it has been widely adopted in solving problems related to complex systems such as transportation, ecosystems, etc. However, in the context of disaster and/or climate resilience assessment, there is limited knowledge on the application of systems thinking (Mavhura, 2017).

The importance of integrating systems thinking in planning sustainable and resilient developments is not only to enhance our understanding of how systems will respond to future climate threats but also plan adaptation pathways. The complex decisions connected with the notion of resilient development are not just ecological, economic, or social, they shall address all three, yet decisions typically short-change one or more (Umberto Pisano, 2012). For instance, adaptation decisions driven by ecological interests may ignore economic or functional constraints, resulting in maladaptive outcomes and hence placing more communities at risk (Juhola et al., 2016). And since resilience is shaped by how systems and external environments interact, evolve, and change over time, we adopt systems thinking in assessing resilience by mapping the cause-and-effect relationship, a.k.a causal relations, as well as circumstantial relations between the system and environment-related factors that may shape the systems' ability to respond to threats posed by sea-level rise.

In this paper, we study the On-Site Wastewater Treatment and Disposal Systems (OSTDS), also known as septic systems. Septic systems treat wastewater coming from individual properties. Initially, wastewater is partially treated in the septic tank, where solid waste rests in the bottom of the tank, and the effluent flows from the septic tank to a drain field. The drain field is a set of perforated pipes that transfer effluent to the ground where it undergoes the final treatment process as it is percolated through unsaturated soils to the groundwater (see Figure 2). In order for septic systems to function properly, the following conditions must be met: (a) soil underneath the drain field is unsaturated, and (b) there shall be a minimum vertical separation distance (VSD) between the bottom of the drain field and wet season high groundwater level. For example, in Florida, this distance ranges from 12 to 42 inches[3] according to the soil percolation characteristics.

With the rising sea levels, septic systems face higher risks of surface and in-land flooding, both of which cause septic system drainfields not to function properly or experience complete failure. Failed septic systems will result in many financial risks for buildings, and their occupants, such as substantial investment in repairs, public health hazards, or degraded property values. In addition to their economic and health risks, environmental risks are of major concern due to contamination of the fresh water resources through introduction of human-caused Nitrogen (N) in partially treated wastewater to groundwater, springs, and coastal waters.

In order to fully capture the factors shaping the response of septic systems to risks due to sea-level rise, hence the systems' resilience, we first start by mapping the risks due to sea-level rise to different modes of failure of septic systems. We visualize our hypothesized Causal Loop Diagram (CLD) from the risk perspective in Figure 2. In this study, our CLD represents the output of all aggregated data, including the minimum requirements for feasible distances dictated by the OSTDS design, siting and management manual published by the U.S. Environmental Protection Agency (EPA 625/1-80-012), and the Florida Administrative Code (rule chapter 64E-6: Standards for OSTDS). In addition to the septic vulnerability report prepared by the Miami-Dade County Department of Regulatory and Economic Resources, in collaboration with Miami-Dade County Water and Sewer Department and Florida Department of Health, as well as other published reports and studies related to septic system vulnerability. We conducted several interviews with representatives from the Miami-Dade Water and Sewer Department (WASD) to validate the scope of our model and the factors.

[3] According to Florida Department of Health administrative standards for onsite sewage treatment and disposal systems.

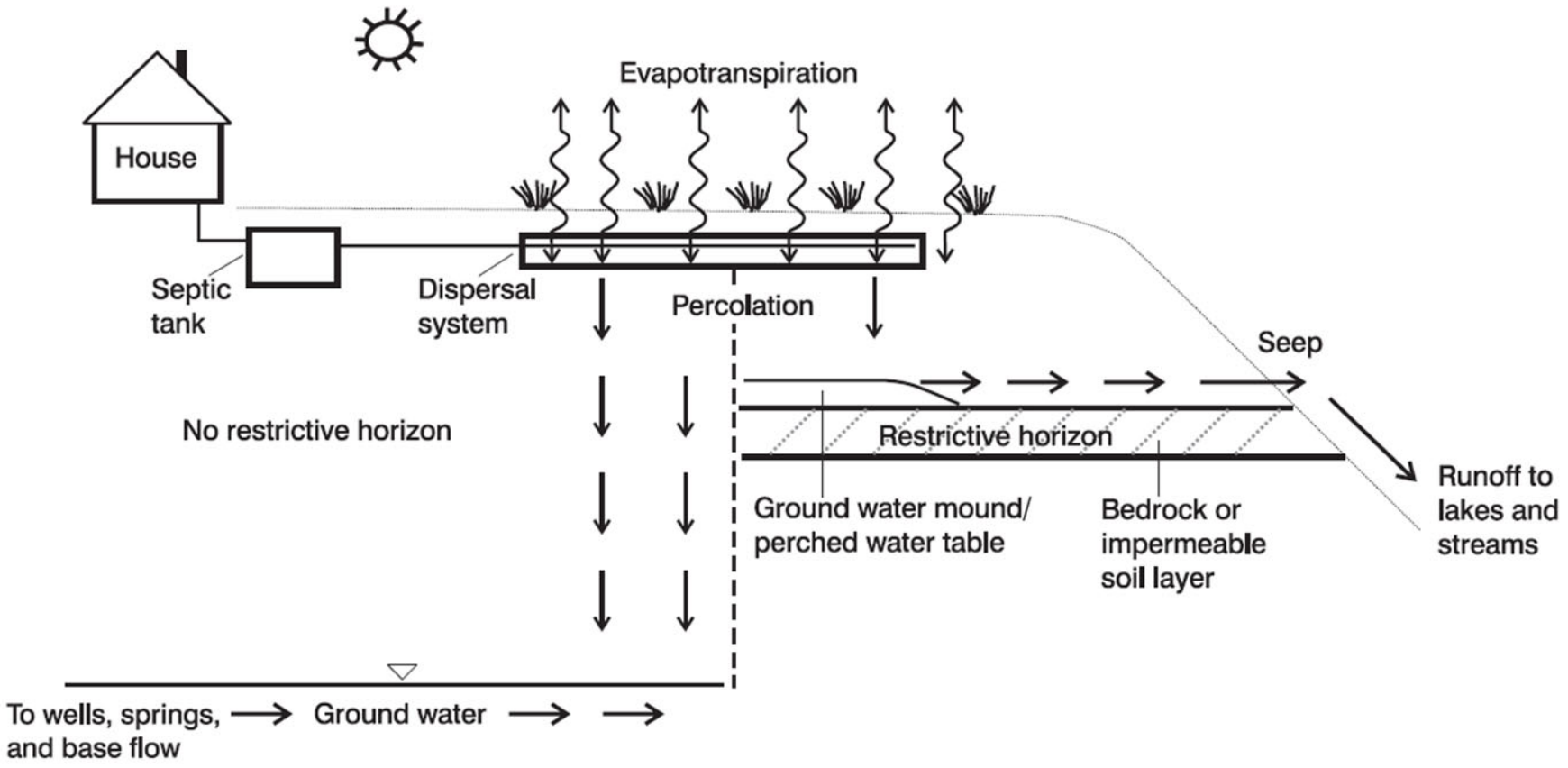

***Figure 2.*** *Operation of conventional OSTDS and flow of discharged Wastewater from property to waterbodies. Source: Venhuizen 1995*

Based on our CLD, as shown in Figure 3, we shortlist 12 factors (attributes) that contribute to the resiliency of septic systems. These are the ratio between daily gallons per day (gpd) and design capacity (percentage of redundant capacity), distance to surface drainage lines (also known as watersheds), distance to high-risk flood zones and the base flood elevation, moisture content in the transfer medium (soils), vertical separation distance, distance to private potable water wellheads, location with respect to wellfield protection zones, distance to surface streams and canals, and the likelihood of groundwater contamination.

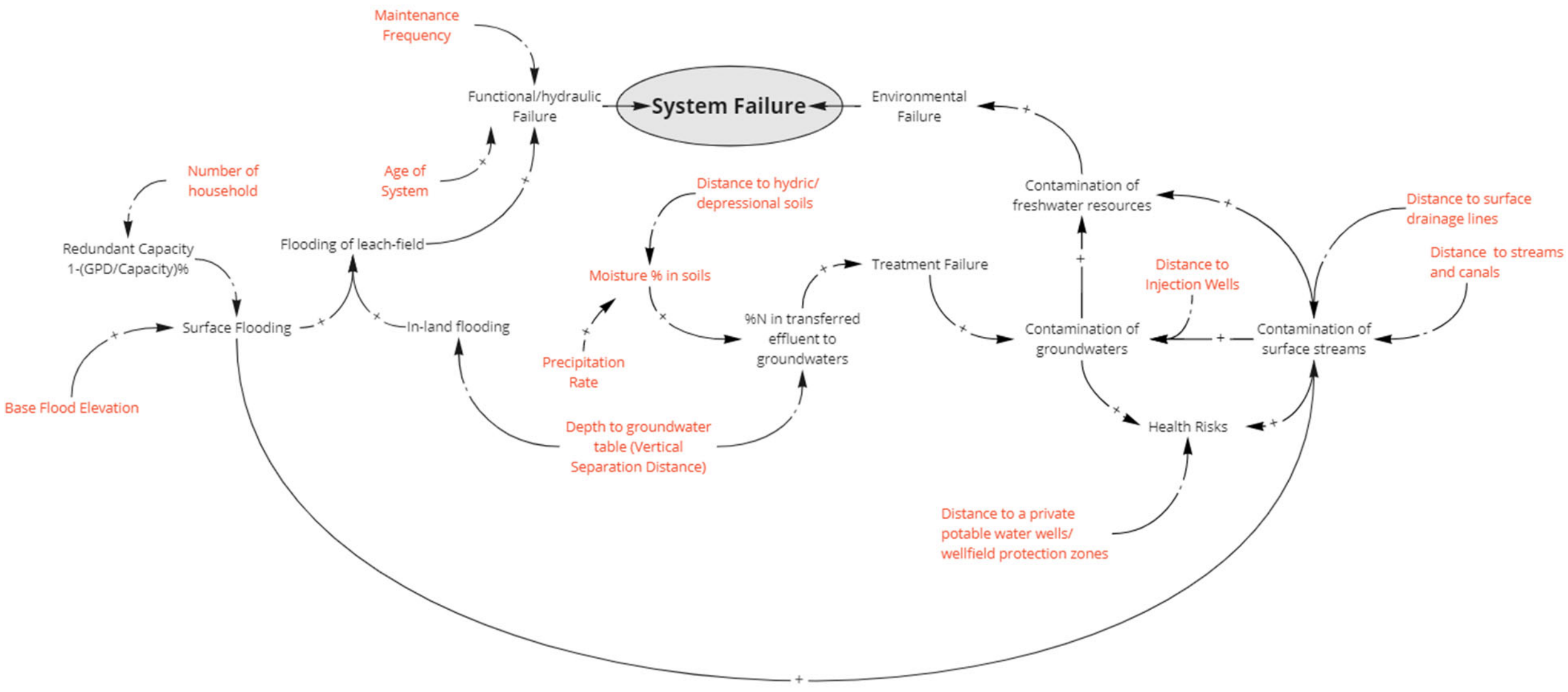

***Figure 3.*** *Causal Loop Diagram for the operability of septic systems from the perspective of risks due to sea-level rise*

By definition, resilience categorizes the system's response to risks into three main phases; prevention (shaped by the system's ability to resist disruptions), damage propagation (the system's ability to adapt to changes while causing minimum damages), and the restoration stage (recovery to original state) (Charani Shandiz et al., 2020). Building on these phases, we categorize the identified factors into three groups, namely, resistivity, adaptability, and ability to recover, as illustrated in Figure 4. It is worth noting that additional factors are considered to shape the ability of the septic systems to recover, yet they do not contribute as a source of risk and as such, not included in the CLD shown in Figure 3. These factors are comprised of distance to existing and/or planned sewer lines, the median household income, and distance to a sewer overflow location. Although these factors do not contribute to the septic systems' vulnerability to anticipated risks, they function as main drivers when it comes to integrating our proposed resiliency metric into adaptation decision making, as discussed later in Section 3.

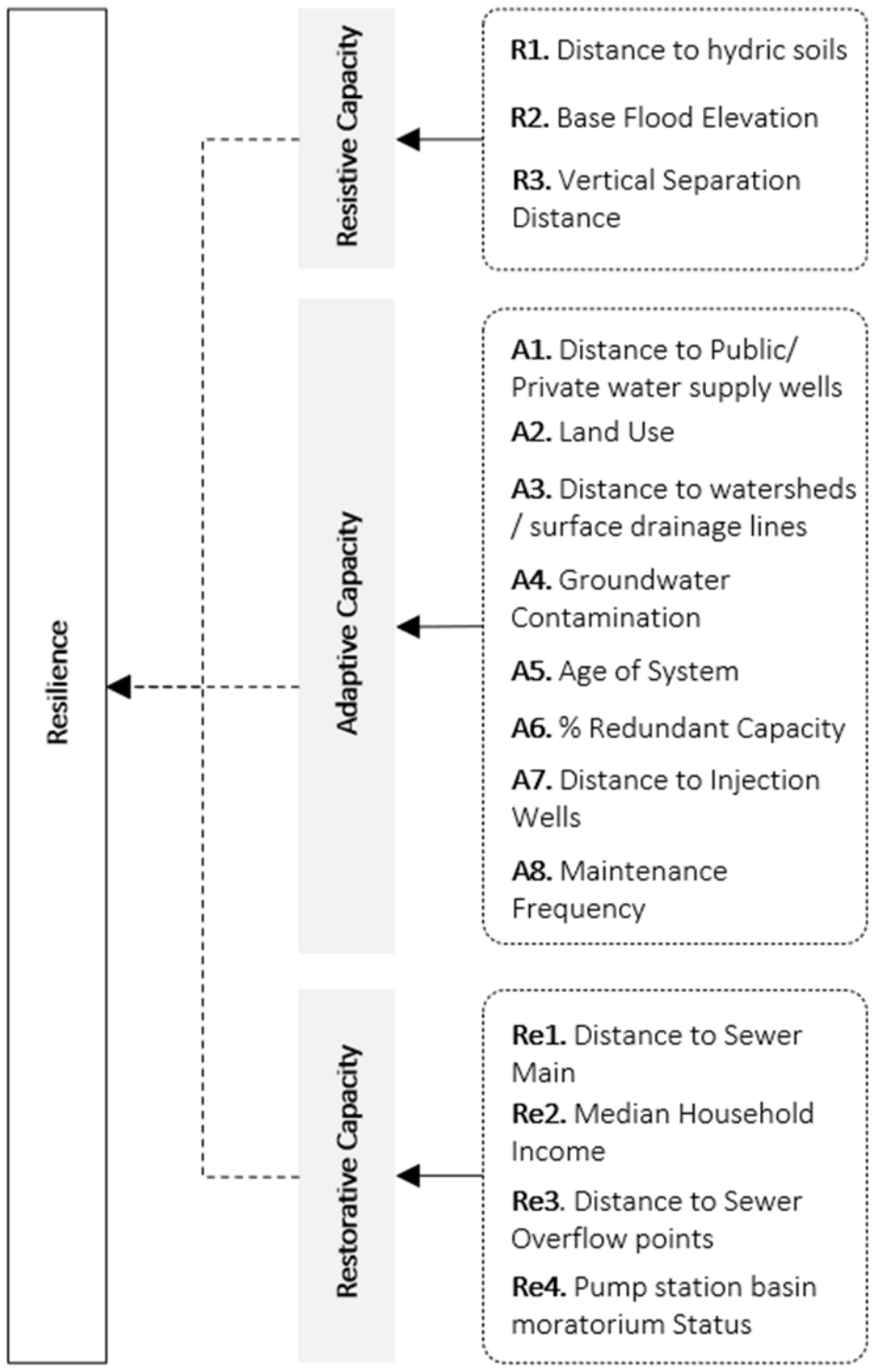

*Figure 4. Categorization of parameters shaping septic system's resilience*

For septic systems, disruptions are either hydraulic or environmental. On one hand, hydraulic disruptions will result in the inoperability of septic systems due to complete surface or inland flooding of the drain field. Surface flooding happens if the system is overloaded, meaning that the daily discharged gallons per day (GPD) exceed the system capacity. This type of failure is not directly related to risks due to sea-level rise under optimum operating conditions and proper maintenance. Also, when the drain field is located within high-risk flood zones, surface flooding of the leach field is very likely to occur. Whereas in-land flooding happens when the vertical separation distance between the bottom of the leach-field and the high season groundwater table approaches zero. On the other hand, environmental disruption occurs due to the conveyance of not-fully treated wastewater to the groundwater as a result of the reduction in the vertical separation distance beyond the minimum required threshold or due to the compromise of the transfer medium by saturation of soils due to flooding or excess rainfall (saturated soils are also known as hydric soils). In the event of hydraulic or environmental disruptions, the septic system will fail to resist failure and hence, factors R1 through R3, as shown in Figure 4, are thought to shape the resistive capacity of the system.

Provided that a septic system fails to resist, we identify some factors that will shape the intensity of the impact and the degree of impact propagation to other critical infrastructures. For instance, a system located very close to or within wellfield protection zones shall be deemed critical as compared to other systems, such that in the event of failure of the septic system not only will groundwater be contaminated, but it is very likely that these polluted waters be drawn into wells that supply drinking water. The same applies to proximity to private wells. According to Florida Department of Health 2020 statistics, nearly 12% of the state population relies on private wells for home drinking water consumption. These private wells are not regulated under the federal Safe Drinking Water Act, as in the case of public water systems, therefore unobserved failure of septic systems will pose health risks to the concerned residents. Also, with industrial waste, there is a larger potential for contaminants and hence land use is a factor to consider in assessing the intensity of impact in the event of failure. Also, environmental risks may be exacerbated if the system at risk is located very close to surface drainage lines (also known as watersheds), these surface drainage lines will transfer untreated wastewater to nearby water bodies and canals propagating both environmental and health risks.

In terms of the system's ability to recover, we consider distance to existing or planned sewer lines, median household income, and distance to sewer overflow locations. The closer the site to the sewer network, the less effort and investments are needed to abandon the septic system at risk and extend the site to the sewer network. However, an overloaded section of the sewer network may experience sewer overflows itself and as such, even though a septic node might be very proximal to an existing sewer line, decision to connect the former to the latter is not always a straightforward option. Therefore, sewer overflow is considered as a factor that shapes recovery ability in the analysis. Also, while households with higher income are more likely to be willing to invest in the recovery of their sites, federal and/or state investments may be needed for low-income areas for recovery and adaptation. As such, the location of the site from the socio-economical perspective is another key component in the overall resilience of systems.

After identifying the various factors contributing to the resilience of septic systems, in the following sub-sections, we discuss how these factors are linked and mapper into a resiliency index and discuss the role of spatial analysis techniques in the context of our proposed resilience assessment approach.

### *2.2. Feature Engineering: Geospatial Analysis*

Geographic Information Systems (GIS) and spatial analysis have been widely used to support disaster risk reduction efforts by deploying spatial analysis models to investigate the sources of risks and identify vulnerability of communities. This type of analysis helps developing and priroritizing actionable plans to reduce community exposure to hazards and increase their resilience by supporting informed planning decisions. In the context of modelling risks that sea level rise poses for OSTDS, Miami-Dade County Office of Resilience has published a study on assessing vulnerability of septic systems to risks due to rising groundwater level and precipitation rates (Elmir 2018).

In making adaptation decisions, it is not only important to assess vulnerability but also resilience since vulnerability often disregards the criticality and impact propagation, an essential need for prioritization. Therefore, in our study, we extend Miami-Dade's septic vulnerability study by integrating the resilience component. Before discussing our proposed resilience index generation methodology, we first present in this section our data preparation efforts and the role of spatial analysis in generating new geospatial dataset of the identified factors shaping resilience of OSTDS. For instance, in order to generate distances between the septic nodes and the main sewer line, we employ proximity analysis and generate the geodesic distances the polygon sewer network data. Figure 5 presents the spatial feature engineering process to transform the available open-access data into a geospatial database compatible with our proposed resilience assessment model.

Our approach is implemented using a case study in Miami Dade County. To carry out the geospatial analysis, we define two sets: the set of active septic sites (*N*) where $N = \{1, 2, \dots, n\}$ and the set of resilience factors (*M*) derived from the CLD process detailed in Section 2.1., where $M = \{1, 2, \dots, m\}$. As mentioned earlier, we identified the factors ($A_1$:$A_8$, $R_1$:$R_3$ and $Re_1$:$Re_4$) shown in Figure 4 to be the most critical system-related attributes that shape resilience in this case study.

The datasets were mostly obtained from the Miami-Dade Open Data Hub, including the existing OSTDS data, parcel data, FEMA flood zones, public and private wellheads, sewer network, sewer overflow points, soil characteristics, waterbodies, etc. The LiDAR-derived Digital Elevation Model (DEM) has a 5ft resolution, and the groundwater raster data has 250 m resolution. Using these datasets, we form an $n \times m$ matrix which we denote by *X*. Here, we let $x_{ij}$ represent the value of the measure for site *i* on factor *j*.

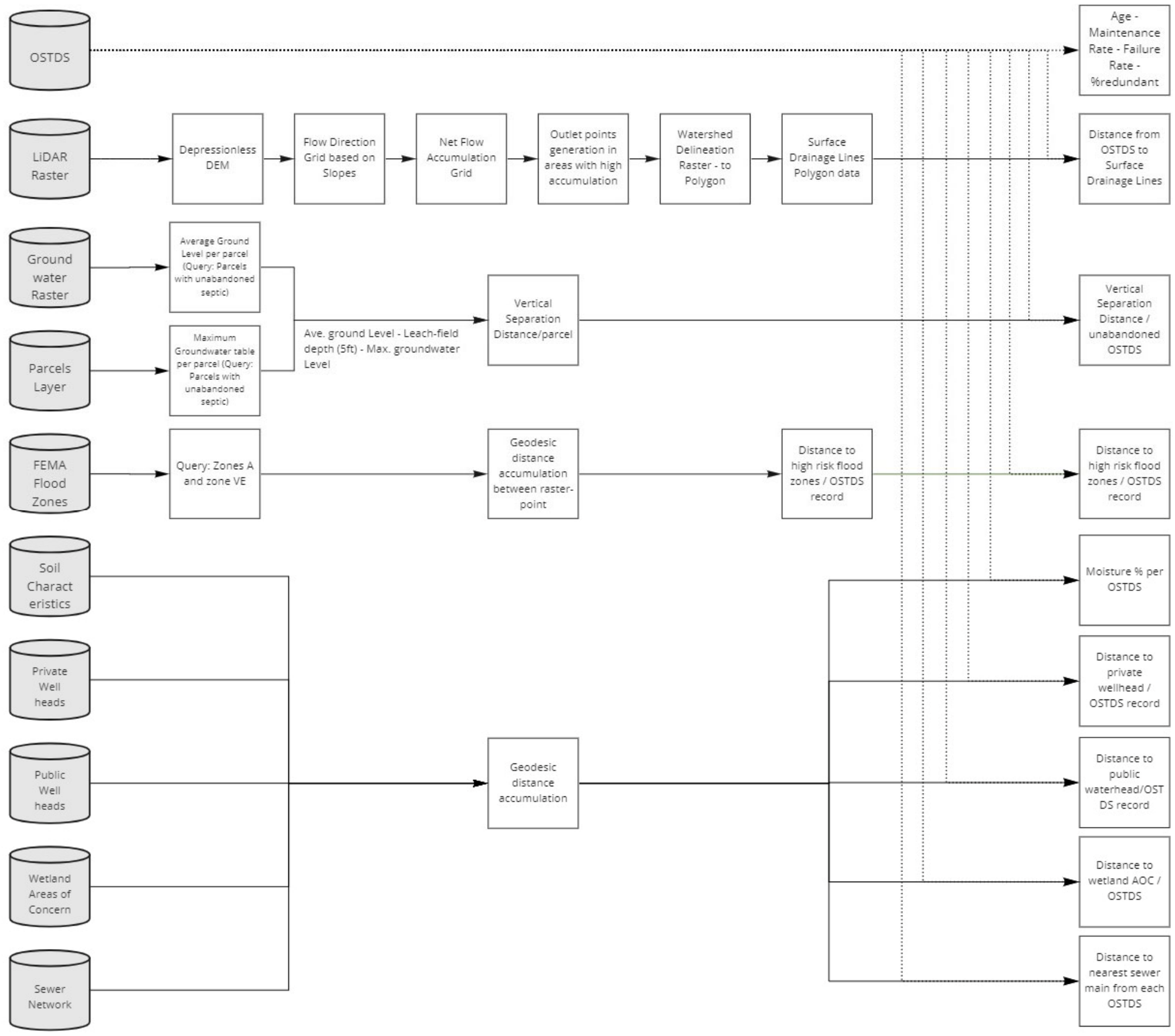

*Figure 5. Process for generating the input geospatial dataset for the resilience assessment model*

As detailed in Figure 5, the input datasets are utilized to generate new features that represent our factors of interest. For instance, to calculate the vertical separation distance (VSD) between the bottom of the drain-field and groundwater table for each septic site, initially, an average ground level is calculated for each parcel with an active septic system. Following the USDA guidelines, we subtract 3-ft for standard drain field depth from the calculated gournd levels. We then compute the maximum groundwater level and subtract again to identify the overall vertical separation distance for each site. Figure 6 shows the raster layers with color coded VSD clusters.

To calculate distances between septic sites and to the nearest surface drainage lines, initially surface drainage lines are populated based on the cell elevations along with their slopes to specify flow direction from each cell to its downslope neighbor(s). Next, drainage lines where surface water runs-off are assembled based on the flow direction. Distances to drainage lines are depicted in Figure 6, where the blue lines represent the drainage lines.

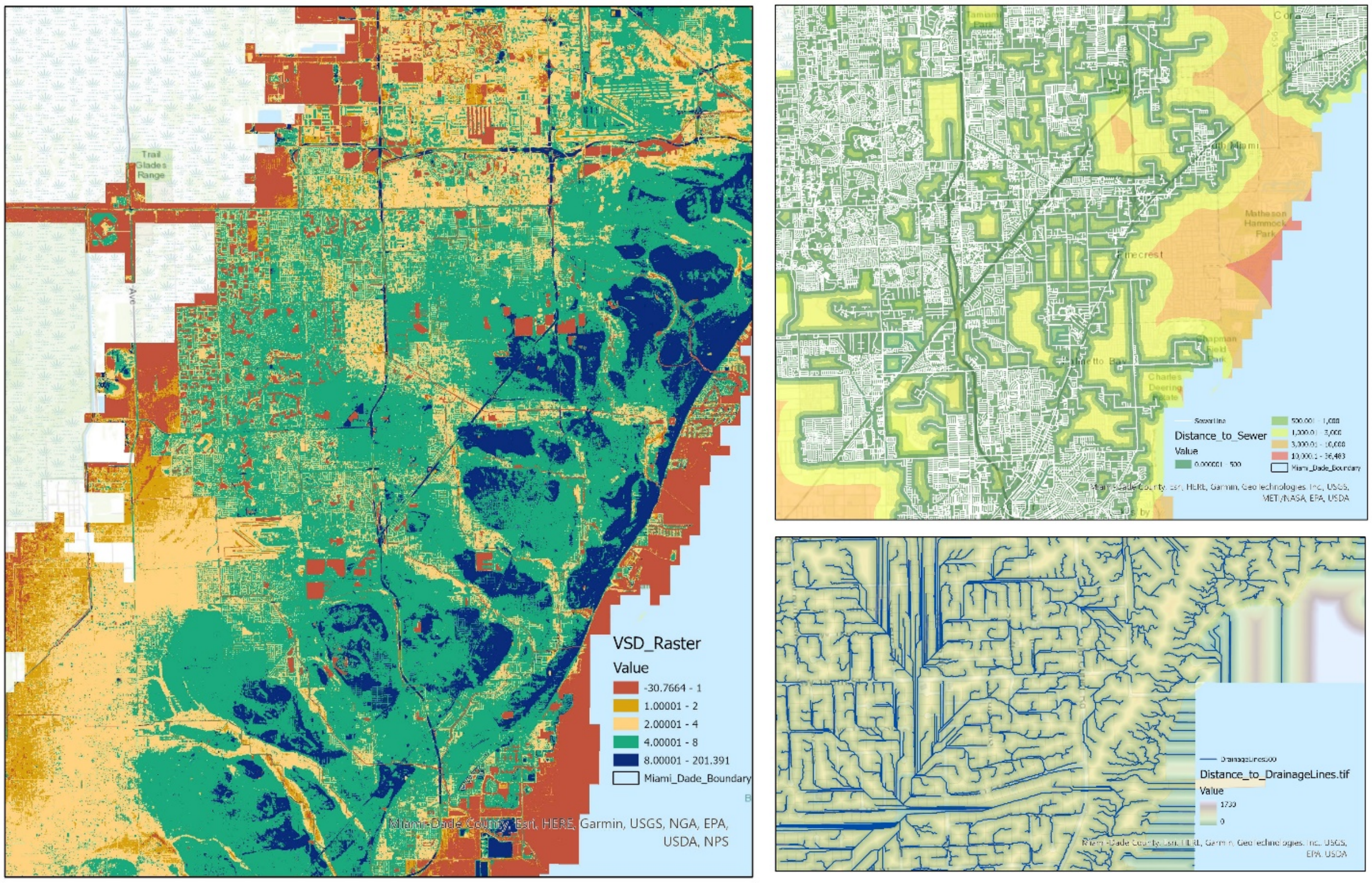

***Figure 6***. *Snapshots from the generated map layers (clockwise from left: a. Vertical Separation Distance, b. distance to sewer lines, and c. distance to surface drainage lines)*

### *2.3. Data Transformation: Fuzzification*

Since it is often difficult to quantify absolute resilience without any existing benchmark with which to validate the calculations (Schneiderbauer, Ehrlich et al. 2006), indicators are typically used to assess relative resilience, either to compare between places or to analyze resilience trends over time (Cutter, Barnes et al. 2008). As there is no preexisting research or paradigm that addresses septic systems' resilience to risks imposed by the sea level rise, we adopt a systems-based method and measure the resilience of the septic sites relative to one another with the goal of informing decisions on prioritizing adaptation actions and assessing their impact on the wastewater treatment system as well as other interrelated infrastructure.

In our proposed methodology, we employ fuzzy logic in handling the imprecision associated with quantifying resilience. As such, rather than dealing with the Boolean scale, we design our resilience index in functional form returning values ranging between 0 and 1, where 0 represents completely not resilient while 1 represents perfectly resilient. This approach enables a continuous function with higher values indicating a higher likelihood of being resilient. The functional form of resilience measure provides two critical advantages. First, such measures can be utilized to order infrastructure nodes in terms of criticality and prioritization. Second, in the context of decision-making, decision variables (i.e., adaptation options) can be mapped into the resulting continuum to assess post-adaptation resilience levels. Consequently, objective functions and constraints can be constructed and incorporated effectively and structurally into decision-making models pertaining to adaptation.

The fuzzification process transforms raw measurements (e.g., VSD) denoted by $x_{ij}$, where $i \in N$ is the index of the septic site and $j \in M$ is the resilience-shaping factor, into scaled and normalized values by applying the fuzzification mapping. This fuzzification process generates a fuzzy set $\tilde{A}_j$ where elements have degrees of membership with respect to factor $j$. Then the set $\tilde{A}_j$ is an array of dual pairs that include raw measures of factor $j$ and their mapping on resilience of septic site $i$. Namely,

$$\tilde{A}_j = \left\{ \left( x_{ij}, \eta_{\tilde{A}_j}(x_{ij}) \right) | i \in N \right\}, \qquad j \in M \tag{1}$$

where $\eta_{\tilde{A}_j}(x_{ij})$ is the output of the fuzzy transformation function that captures the degree of membership of septic site $i \in N$ in fuzzy set $\tilde{A}_j$ for a given measurement of $x_{ij}$ and returns a value in $[0,1]$.

Multiple fuzzy transformation functions are considered for varying contexts. For instance, we use the Sigmoid function as presented in (2) for transforming the VSD data, where larger input values are more likely to be members of the higher resilience set, as illustrated in Figure 7. In this case, the midpoint is set at 3-ft, which corresponds to membership likelihood of 0.5. Below (*resp.* above) this value, the likelihood of being resilient decreases (*resp.* increases) monotonically.

$$\eta_{\tilde{A}_j}(x_{ij}) = \frac{1}{1 + \left( \frac{x_{ij}}{f_{2j}} \right)^{-f_{1j}}} \tag{2}$$

In (2), $x_{ij}$ denotes the input value of parameter $j$ for site $i$, $f_{1j}$ is the parameter to control the shape of the curve, and $f_{2j}$ is the reference value (e.g. 3 ft for the VSD case). In the context of OSTDS, reference values and/or boundaries can be determined based on the recommendations dictated by the OSTDS design, siting and management manuals published by the U.S. Environmental Protection Agency (EPA 625/1-80-012), and the Florida Administrative Code (rule chapter 64E-6: Standards for OSTDS). In the absence of regulated feasible distances, such as in the cases of distance to surface drainage lines and sewer overflow, the reference values are selected based on the median values in the dataset. In such cases, we assess the resilience of the systems in relevance to one another.

In cases where smaller values imply higher resilience, such as distance to sewer network, we empoy the inverted logistic S- function, represented as follows:

$$\eta_{\tilde{A}_j}(x_{ij}) = \frac{1}{1 + \left(\frac{x_{ij}}{f_{2j}}\right)^{f_{1j}}} \quad (3)$$

In this case, this function decreases from *1* and converges to *0* as $x_{ij}$ increases, following an inverse S-shape, as illustrated in Figure 7. Since there are no requirements for how proximal a site has to be from the sewer network, $f_{2j}$ is chosen as the median value in the dataset as mentioned earlier.

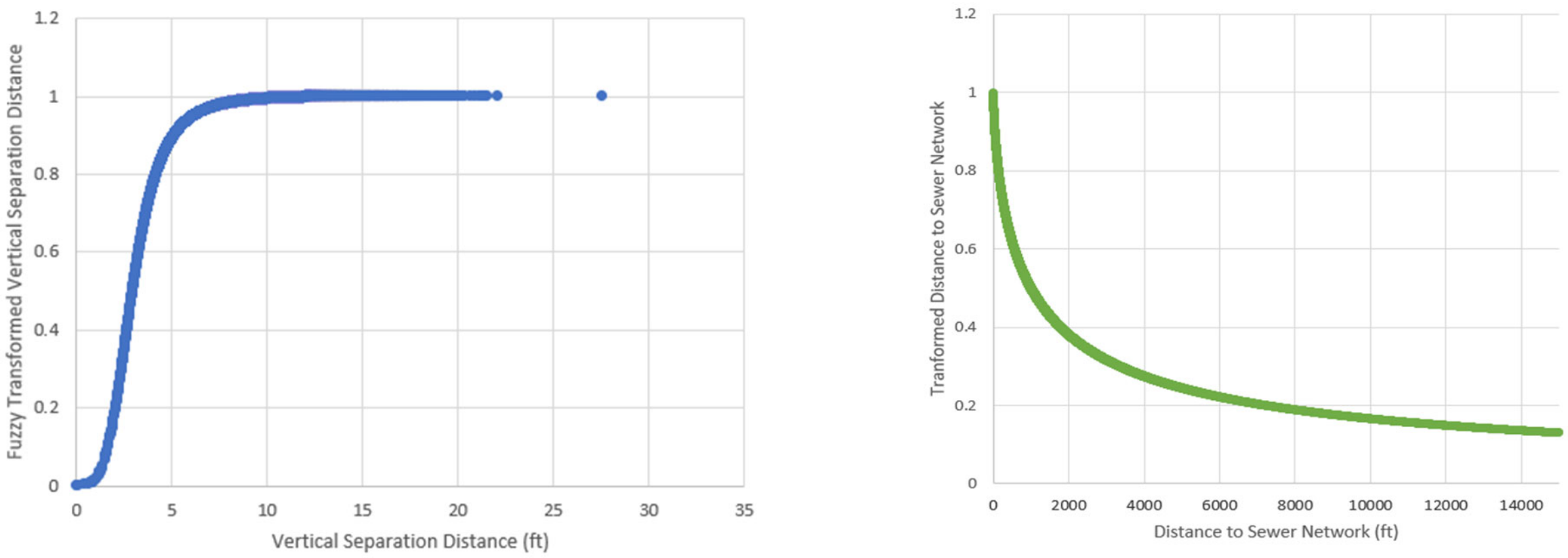

***Figure 7.*** *Examples of the performed fuzzy transformations. The input value is shown on the x-axis and the transformed value is shown on the y-axis*

The shape of the curve can be adjusted to reflect the relative importance of the different factors shaping the resilience of the entire system. For instance, a septic system located at 100 ft from wetland areas of concern (areas with saturated and decompressed soils) is considered to be more resilient compared to another system located at 100 ft from a potable water wellhead, provided all other factors remain the same. This is because although the system is very close to wetland areas of concern, it still meets the required operating conditions as long as the soil underneath the drain field is suitable for treatment, whereas the latter does not meet the minimum required feasible distances to wellheads (200 ft from public wells). These thresholds are captured by the shape and mid-point of the transformation functions. To illustrate, Figure 8 compares the transformation function for three factors: distance to wetland areas of concern, distance to surface canals and streams, and distance to potable water wellheads.

In cases where relative transformations cannot be conducted, such as transforming the vertical separation distance, where no other factors are referenced to the vertical datum, we account for both current and anticipated future sea-level scenarios and the associated rise in the groundwater table so as to ensure that the factor value converges to 1 implying perfect resilience or zero-risk condition. According to the IPCC 6th Assessment Report, under the intermediate greenhouse gas emission scenarios, global sea levels are projected to rise by 0.56 m ± 0.2 (1.837 ft ± 0.656 ) by 2100. Moreover, according USGS and other studies that assess SLR-induced groundwater rise, such as Knott, Jacobs, et al. (2019), the projected mean groundwater rise relative to sea-level rise is expected to be 31 to 35%[4]. This means that by 2100, groundwater table is expected to increase by 0.7ft. Under such scenarios, systems with vertical separation distances being greater than or equal to 5 ft will function properly, provided that all other conditions are ideal. Based on this, the vertical separation distance transformation is adjusted in a way that it converges to 1 around 5-6 ft levels as illustrated in Figure 7.

[4] Although the projected mean groundwater rise is subject to change under various conditions, including location, land topology, surface-water hydrology, distance inland further from the shoreline, etc., we assume the same relationship applies for our case study until further research emerges in this context.

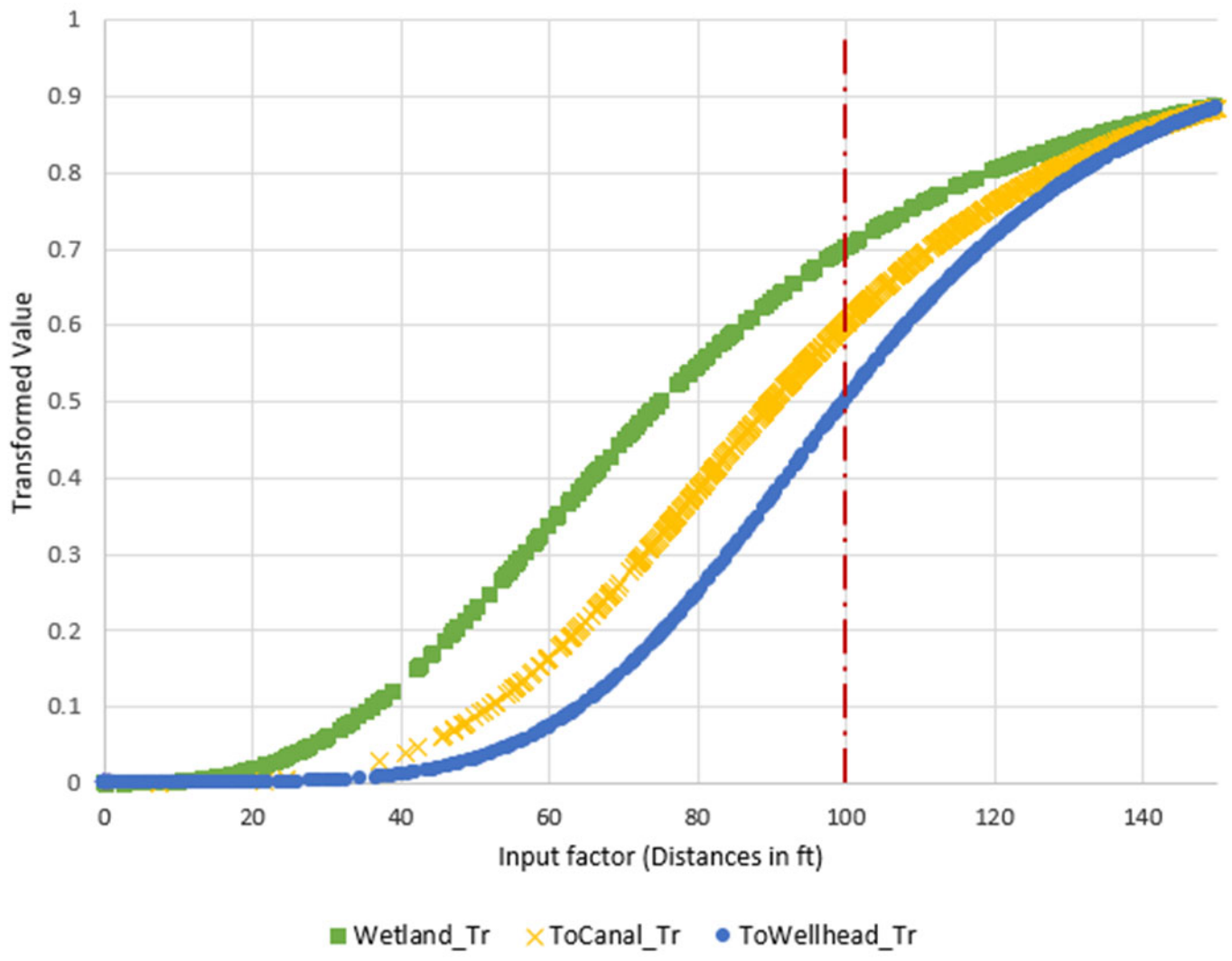

***Figure 8.*** *Sigmoid transformation functions for distance to wetlands, canals, and distance to potable water wellheads*

In addition to the Sigmoid and the inverted logistic S-transformations, in the absence of reference points, fuzzy grade and/or inverse grade transformation functions can also be used, as in distance to the sewer overflow points. In this case, the transformation is linear and a function of minimum and maximum acceptable values. When resilience is perceived as an increasing function of the parameter measure, we can use the following function:

$$\eta_{\tilde{A}_j}(x_{ij}) = \begin{cases} 0; & x_{ij} \leq x_j^{min} \\ \dfrac{x_{ij} - x_j^{min}}{x_j^{max} - x_j^{min}}; & x_j^{min} < x_{ij} < x_j^{max} \\ 1; & x_{ij} \geq x_j^{max} \end{cases} \tag{4}$$

In the opposite case we use the following:

$$\eta_{\tilde{A}_j}(x_{ij}) = \begin{cases} 0; & x_{ij} \leq x_j^{min} \\ \dfrac{x_j^{max} - x_{ij}}{x_j^{max} - x_j^{min}}; & x_j^{min} < x_{ij} < x_j^{max} \\ 1; & x_{ij} \geq x_j^{max} \end{cases} \tag{5}$$

For example, in transforming distance to sewer overflow, any system located within 200 ft from an existing sewer line can be assigned a value of 1 (i.e., fully resilient), and any site located more than 1000 ft from an existing sewer line can be assigned a value of 0 (i.e., completely not resilient).

In general, the nonlinear transformations governed by functions in (2) and (3) result in more precise measures compared to the linear transformations defined by (4) and (5), especially when reference points derived from design manuals can be obtained in the former cases. In addition, the functions used in the former approach have innate bounds of 0 and 1 and as such, they provide higher degrees of freedom within this range. The linear transformation requires minimum and maximum bounds on the factor measures whose identification may result in a nontrivial - and possibly a subjective - process of setting limits for the resiliency categories. However, using this transformation will still be advantageous due to its linear structure, especially in the context of large-scale decision models, where computational complexity in finding solutions is a concern.

### *2.4. The Resultant Resilience Indicator: CRI-DS*

In order to overcome the challenges posed by earlier proposed strategies as discussed in section 1, we build on the theories of reliability and systems engineering and propose a systematic parameter aggregation strategy for computing the overall resilience index for each system. Systems engineering views the systems as complex structures composed of connected multiple elements and modules where the mutual arrangements of the individual elements influence the resultant system reliability. According to earlier resilience theories, the perception of resilience is akin to reliability (robustness) associated with elasticity of the system that is reflected in its ability to recover (Shi, Watanabe et al., 2018). Since factors shaping ability to recover are included in our model, we use resilience interchangeably with reliability in our discussion.

Based on this rational, we suppose that a septic system embodies various components arranged together and collectively shape its resilience. We contend that for the septic system to be perfectly resilient to risks due to sea-level rise, all resilience-shaping factors must achieve a minimum degree of marginal resilience level. In our context, this means that all individual metrics obtained by the fuzzy transformation functions introduced in the previous subsection must converge to 1. In this respect, in case some factors do not possess a high degree of isolated resilience, we still need a mapping to capture the overall resilience index for a given site. Our approach proposes a methodology to tackle this challenge as explained in what follows.

The proposed mapping is founded on two fundamental observations: resilience-shaping factors are not equally critical and as such, an effective mechanism is needed to capture the criticality of these categories for combining them into a single resilience metric. With this understanding, our methodology links the resilience-shaping factors to each other based on their roles in the overall functionality and response of the system. This is achieved with the help of a systematic block diagram as illustrated in Figure 9. The block diagram is utilized to formulate a configuration where the factors are related to each other in serial or parallel fashion resulting in an acyclic graph. The diagram is constructed in a way that ensures the inclusion of all factors defining the system's ability to resist, adapt and recover from disruptions. A resilient system is one that is able to resist disruptions by continually functioning at its desired performance level and also able to respond effectively following a disruption. While the degree of resistivity shapes the functionality, both adaptability and ability to recover relate to response. Therefore, a system that is not able to adapt but can still recover effectively is not fully resilient due to impact propagation. In that respect, factors shaping response - adaptability and recovery – must be connected in series. Consequently, the system needs to achieve a high degree of isolated effect on resilience in both categories so that the system's response to discurption can be effective. However, functionality and response are connected in parallel to reflect the perception that if a system fails to resist, and eventually fails to function, it can still maintain a degree of resilience depending on its ability to respond.

It is worth noting that since adaptability accounts for impact propagation, such as groundwater contamination, if the vertical separation distance is not sufficient and/or the soils are saturated, partially treated waters will be transferred to the groundwater eventually, and as a consequence, the system's ability to adapt will be impacted, reducing the overall system's resilience. Factors associated with blocks, which are listed in Figure 4, are connected in series within their respective blocks to shape the overall resilience structure of these three categories.

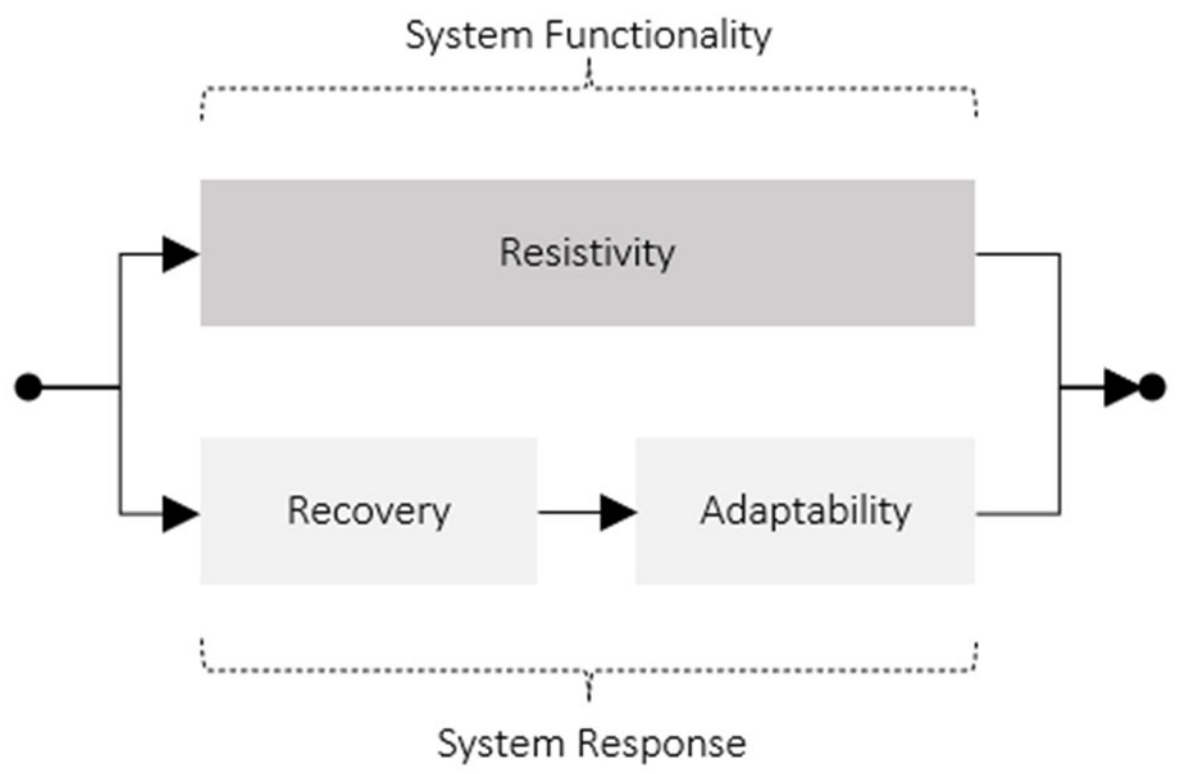

*Figure 9. System block diagram for factors shaping resilience*

Building on this reasoning and the hypothesized casual relationships, we introduce the following conjectured formulation to measure the resultant resilience index. This formulation is derived by gradual simplification and the application of the convoluted block reliability formulas. We can write the specific resilience index formula $CRIDS_i$ for site $i$ as follows:

$$CRIDS_i = 1 - [1 - \prod_{j \in J} \eta_{\tilde{A}_j}][1 - (\prod_{k \in K} \eta_{\tilde{A}_k})(\prod_{z \in Z} \eta_{\tilde{A}_z})] \qquad (6)$$

where $J = (R_1, R_2, R_3)$, $K = (A_1, A_2, A_3, A_4, A_5, A_6, A_7, A_8\ ), and\ Z = (Re_1, Re_2, Re_3, Re_4)$ represent sets of factors shaping resistivity, adaptability and ability to recover, respectively.

For illustration purposes, we selected three septic systems located in Miam-Dade County to investigate how various conditions impact resiliency of the septic sites. As shown in Table 1, for the system identified as AP497567, the vertical separation distance is 20 ft, distance to wetland areas of concern is around 9435 ft etc. For this system, although adaptability is very poor (nearly zero), overall system resiliency is considerably high. This is because no impact propagation is anticipated since the system is very far from wellheads, surface canals, and no groundwater contamination is expected due to the large vertical separation distance and good soil conditions. The only risk in this case is due to the site's proximity to surface drainage lines, which may transfer sewage water tp these lines in the event of failure. However, given its very high ability to resist, risks are negligible and hence overall system resilience is relatively high.

In another example, although the resistivity of the site identified as AP1584897 is low, the overall system's resilience is moderately high, due to the average likelihood of impact propagation, as reflected in the adaptability metric, in addition to its medium ability to recover. Another case is concerned with medium vertical separation distance, represented by site AP1204641, where the vertical separation distance is around 3 ft. In this case, although the system is considered to meet the design requirements and the soil conditions are proper ensuring full treatment of the effluent as it flows through soil to the groundwater, the overall resiliency is low. This is due to the effect of impact propagation. This system is located within a high-risk flood zone and very close to surface drainage lines, which will accelerate transferring wastewater to streams and canals with likely occurance of surface flooding.

According to the Department of Health OSTDS data obtained from Miami-Dade open data hub[5], the County has a total of 107,526 active septic sites. In our analysis considering current sea-levels and flood-risk zoning, we have identified that nearly 32% of these systems carry a resiliency index below 0.5 and around 8% of them have resiliency index less than 0.1, as illustrated in Figure 10. These results indicate a need for developing and implementing adaptation actions in near future.

***Table 1.*** *The calculated Resilience index for three different septic systems located in Miami-Dade County subject to different operating environments*

*Original Values*

| APNO† | VerticalSepDist | BaseFloodElev | Dist.Wetland | | Dist.Wellhead | Dist.Canal | Dist.SDrainage | System_Age | Dist.Sewer | Dist.Overflow |
|---|---|---|---|---|---|---|---|---|---|---|
| AP497567 | 20.1515 | 0 | 9435.767 | | 867.3902 | 1094.696 | 0.4827153 | 122.2016 | 413.0418 | 1848.942 |

*Transformed Values*

| VSD_Tr | BFE_Tr | Wetland_Tr | GWContam* | ToWellhead_Tr | ToCanal_Tr | ToDrainage_Tr | Age_Tr | ToSewer_Tr | Overflow_Tr |
|---|---|---|---|---|---|---|---|---|---|
| 0.918749501 | 1 | 0.999999498 | 0.91874904 | 0.999979633 | 0.999954315 | 1.04013E-09 | 0.05 | 0.711518429 | 0.876710452 |

| Resistivity | Adaptability | Recovery |
|---|---|---|
| 0.9187 | 4.77779E-11 | 0.6238 |

**Resiliency Index = 0.9187**

*Original Values*

| APNO | VerticalSepDist | BaseFloodElev | Dist.Wetland | | Dist.Wellhead | Dist.Canal | Dist.SDrainage | System_Age | Dist.Sewer | Dist.Overflow |
|---|---|---|---|---|---|---|---|---|---|---|
| AP1584897 | 7.60257 | 10 | 7622.325 | | 369.1966 | 3400.174 | 120.9102 | 1.370917 | 794.9039 | 1396.808 |

*Transformed Values*

| VSD_Tr | BFE_Tr | Wetland_Tr | GWContam* | ToWellhead_Tr | ToCanal_Tr | ToDrainage_Tr | Age_Tr | ToSewer_Tr | Overflow_Tr |
|---|---|---|---|---|---|---|---|---|---|
| 0.7238 | 0.309 | 0.999 | 0.7238 | 0.9985 | 1 | 0.8037 | 1 | 0.6093 | 0.8236 |

| Resistivity | Adaptability | Recovery |
|---|---|---|
| 0.2234 | 0.5808 | 0.5018 |

**Resiliency Index = 0.4499**

*Original Values*

| APNO | VerticalSepDist | BaseFloodElev | Dist.Wetland | | Dist.Wellhead | Dist.Canal | Dist.SDrainage | System_Age | Dist.Sewer | Dist.Overflow |
|---|---|---|---|---|---|---|---|---|---|---|
| AP1204641 | 3.12614 | 9 | 1225.942 | | 588.224 | 3456.406 | 61.76985 | 6.483218 | 448.6923 | 452.2486 |

*Transformed Values*

| VSD_Tr | BFE_Tr | Wetland_Tr | GWContam* | ToWellhead_Tr | ToCanal_Tr | ToDrainage_Tr | Age_Tr | ToSewer_Tr | Overflow_Tr |
|---|---|---|---|---|---|---|---|---|---|
| 0.4086 | 0.3204 | 0.9998 | 0.4085 | 0.9999 | 1 | 0.2181 | 0.8 | 0.6995 | 0.4624 |

| Resistivity | Adaptability | Recovery |
|---|---|---|
| 0.130877173 | 0.071257109 | 0.3235 |

**Resiliency Index = 0.1509**

*Groundwater contamination is calculated based on vertical separation distance and distance to wetland areas of concern
†APNO is the septic system application number, it is unique to each system.
All distance measures are in US ft

[5] https://gis-dc.opendata.arcgis.com/datasets/4f036c78c34c41ec99e0e1d955a9c35c/explore

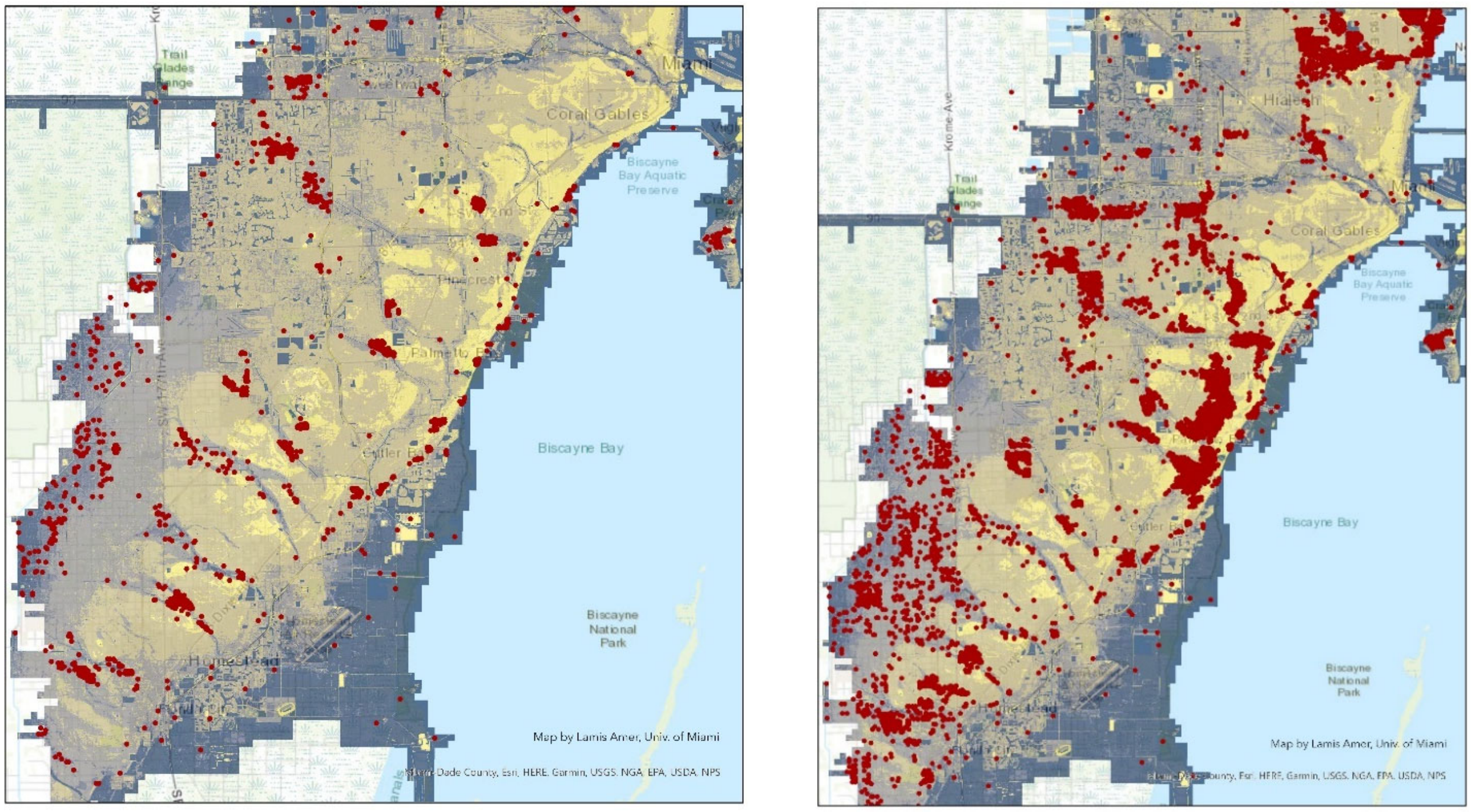

***Figure 10.*** *(left) Septic Systems with Resilience Index less than 0.1 (in red) and (right) septic systems with Resilient Index less than 0.5 (in red)*

## 3. Integration into Decision Making

As emphasized earlier, our work is primarily motivated by developing functional resilience metrics that can be mapped to decision variables and incorporated into the objectives and constraints of decision models. In this section, we present how the proposed CRI-DS can help us achieve such integration. The way the decision variables are linked with the resilience metrics depends on the scope of the former one. If the decision variable is directly associated with the factor measurements, we can map them into resilience metrics directly by replacing them with $x_{ij}$ in the transformation functions discussed in the previous section. An example is the vertical separation distance between the bottom of drain field and the wet season groundwater level in the context of septic systems. Basically, we can set such measures as decision variables and thus, control the resilience metrics via the transformation functions. Clearly, in this case, the transformation functions must be directly incorporated into the objectives and constraints in the decision model.

In case the decision variable is not a one-to-one mapping with a resilience metric, incorporating its impact on the overall resilience index CRI-DS is not a straightforward process. In that case, we must capture the scope and extend of the impact of this decision variable on each resilience-shaping factor. An example in the context of our case study is the decision whether a site with septic system should be connected to main sewer line as part of adaptation actions. Clearly, connecting the site to the sewer line will significantly affect almost all factors and eventually the resilience of the overall system. We employ causal loop diagrams once again to establish the relation between the adaptation decision options and the resilience metrics. The CLD presented in Figure 11 demonstrates the mapping between decisions and the factors shaping the overall system resilience by using our case study as an example. In this example, the CLD links the sewer extension decision to the site resilience index across resiliency-shaping factors. The CLD diagrams are instrumental in helping the analysts link the critical models and assess the degree of change in a system's resilience as a consequence a made decision (Li, Kou et al. 2020, Zhang, Tan et al. 2021).

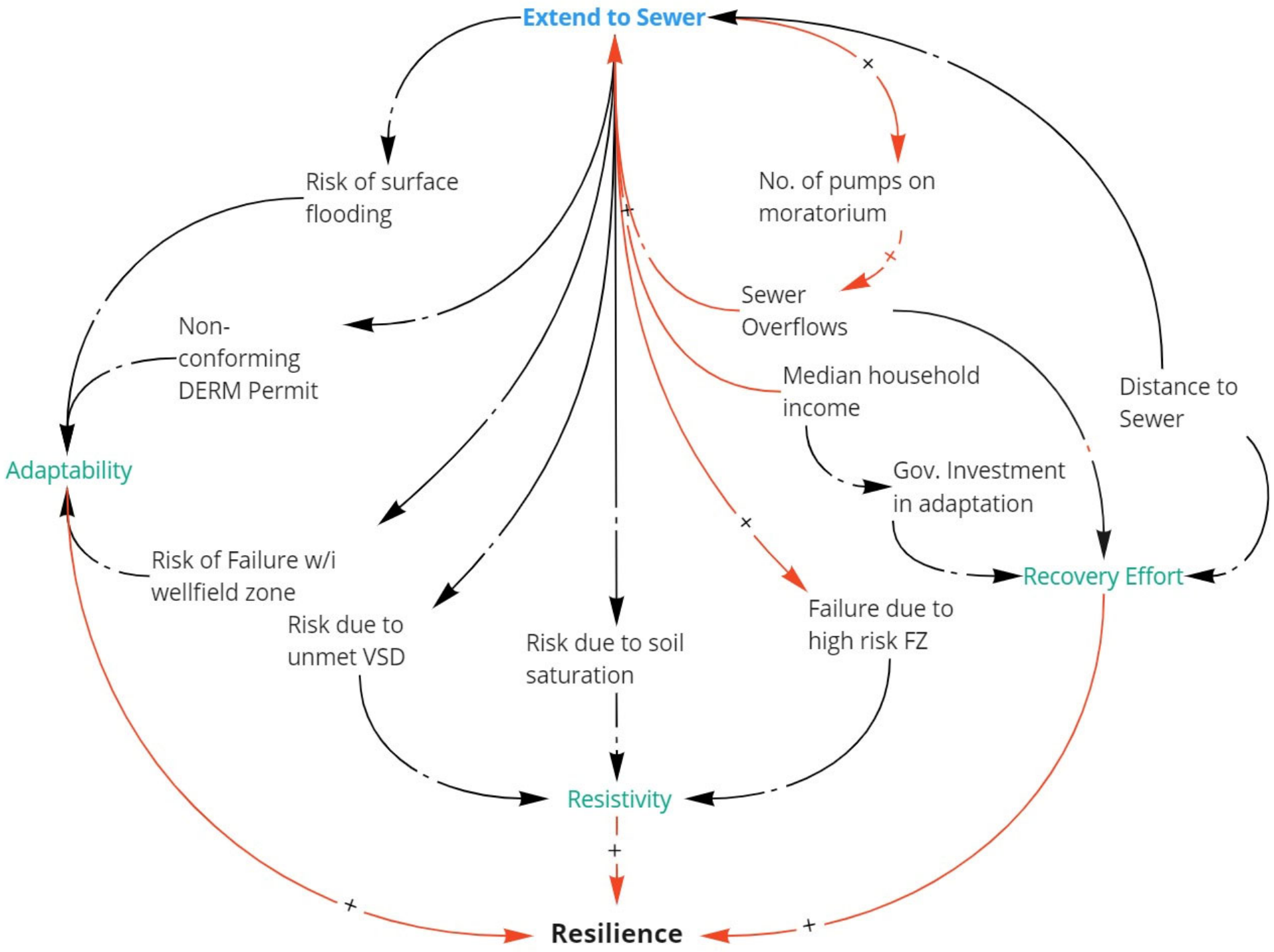

***Figure 11.*** *Causal Loop Diagram from the perspective of adaptation - For illustration: example on Sewer Extension Decision*

Consider the following adaptation decision problem for the septic system. For each septic site in the system, we have the following adaptation options: connecting the site to the existing main server line, connecting the site to a micro treatment plant (also known as community sewer network), elevating (mounding) the drain-field to the minimum VSD level required for full resilience (e.g. 5 ft), installing an onsite advanced treatment system, and do nothing. Let $y_{il}$ denote the decision for site *i* and adaptation option $l$, where $l$ takes values between 1 to 5 in respective to the aforementioned list of adaptation options. Clearly, in this case, the decision variables are binary across all options and they influence the functional measures, i.e., $x_{ij}$, indirectly. As such there is a mapping between the decision variables and the raw measures. Obviously, selecting the do-nothing option should not lead to any changes in the raw meaures and as such, the CRI-DS values calculated by using (6). We note that in this context exactly one of the decision variables must take the value of 1. As am example, when $y_{il} = 1$, the resilience index formula in (6) must be rewritten as the following for a given septic site *i*:

$$CRIDS_{i1} = 1 - [1 - \prod_{j=R_3} \eta_{\tilde{A}_j}][1 - (\prod_{k=A_2} \eta_{\tilde{A}_k})(\prod_{z \in Z/\{Re1\}} \eta_{\tilde{A}_z})] \qquad (7)$$

Based on the relationships identified from the CLD in Figure 11 (arcs with "+" connections), after abandoning the existing septic system and extending it to sewer network, only the following factors would influence the resilience of the new site: $R_3$ (distance to high-risk flood zone), $A_2$ (land use), $Re_2$ (household income), $Re_3$ (distance to sewer overflow) and $Re_4$ (pump station basin moratorium status). All other resilience measures converge to 1. In other words, extension to the sewer network completely mitigates some of the factors such as the vertical separation distance, saturated soils, etc. and hence, these factors no longer constitute threats to the sewer network.

Similarly, by considering elevating the drain-field (mound septic system), the factors; $R_3$ (the vertical separation distance), $A_4$ (groundwater contamination), and $A_5$ (system age) no longer constitute threats to the system, since a new leach-field is to be installed at a higher elevation. The resilience index formula in this case shall be reduced to:

$$CRIDS_{i2} = 1 - [1 - \prod_{j \in J \setminus \{R_3\}} \eta_{\tilde{A}_j}][1 - (\prod_{k \in K \setminus \{A_4, A_5\}} \eta_{\tilde{A}_k})(\prod_{z \in Z} \eta_{\tilde{A}_z})] \quad (8)$$

The resilience index formula for choosing community treatment network follows (7) and the resilience index formula for adopting onsite treatment will be similar to (8), because their influence on improving resilience for a given site is nearly the same as extension to sewer network and the mounding, respectively. Surveys of alternative adaptation options for community based micro treatment plants and onsite treatment technologies are provided by Weiss, Eveborn et al. (2008), Molinos-Senante, Garrido-Baserba et al. (2012), and Abdalla, Rahmat-Ullah et al. (2021). In these cases, the trade-off when comparing between these decisions depends on the cost-benefit trade-off and the feasibility determined by constraints of the decision-making model. For instance, extension to sewer network decision cannot be made when the pump station basin is on moratorium status. Also, mound systems cannot be considered when the vertical separation distance is less than 1 ft.

Typically, a decision model in this context have at least two conflicting goals: mazimizing resilience and minimizing cost. Depending on the perseptions of the decision maker, the objectives and the constraints of the model can be set up in numerios ways. One practical approach to deal with multiple objectives is to minimize the total cost under a constraint set that stipulate minimum resilience requirements for septic sites. As an alternative, one can use a budget constraint and maximize the overall resilience. Using the former modeling choice as an example, we can build a general framework as demonstrated by the following integer programming model:

$$Min\, Z = \sum_{i \in N} \sum_{l \in \{1..5\}} c_{il} y_{il} \quad (9)$$

*s.t.*

$$\sum_{l \in \{1..5\}} y_{il} CRIDS_{il} \geq b_i \qquad \forall\, i \in N \quad (10)$$

$$\sum_{l \in \{1..5\}} y_{il} = 1 \qquad \forall\, i \in N \quad (11)$$

$$y_{il} \in (0,1) \qquad \forall\, i \in N, l \in \{1..5\} \quad (12)$$

In this general formulation, $c_{il}$ denote the cost of choosing adaptation option *l* for site *I* and $b_i$ the minimum resilience level required for the septic site *i*. Constraint (10) ensures that the resulting resilience after adaptation for site $i$ can be no less than the prespecified threshold set for that site. Constraint (11) ensures that exactly one adaptation option is chosen for each site. We recall that do-nothing is included as an option in this case. We note that this formulation is for illustration purposes, and thus, it neglects any possible operational constraints associated with each adaptation decision or constraints that eliminate certain adaptation options that are not technically feasible for certain sites. As mentioned earlier, the left hand side of Constraint (10) can replace the objective function with a maximization goal and it can be replaced by a budget constraint that limit the investment amount. In both cases, a frontier of solutions with non-dominated outcomes can be obtained, which can help decision makers evaluate the most cost efficient plans that achieve preset resilience goals or best resilience outcomes under preset budget levels.

## 4. Conclusions

The threats of damage to many infrastructure systems will continue to increase over time with the rising sea levels. To tackle these threats, policy makers need holistic tools developed based on scientific knowledge that will help them establish policies that are effective in improving resiliency and mitigating effects of the climate change, prioritizing their adaptation portfolios, and monitoring progress. In this paper, we first provide a systematic analysis of what factors constitute the resilience of critical decentralized infrastructure, with a particular focus on on-site wastewater treatment and disposal systems (OSTDS). Then by means of integrating spatial analytics, fuzzy transformation, and concepts of reliability engineering we propose a quantitative novel composite measure for resilience for decentralized infrastructure systems (CRI-DS). The proposed metric integrates system characteristics, as well as various environment and social-related factors that shape the overall system's ability to resist to, adapt to, and recover from disruptions, and hence its resilience. Because our proposed resilience index is a multi-dimensional measure, it can serve as a practical tool for decision-making through mapping the relationships between adaptation decisions and the factors constituting the resilience index formulation.

In addition, we propose a general framework for integrating the CRI-DS metric into an adaptation decision-making modeling using mathematical programming. The introduced framework can effectively map decision variables regarding adaptation strategies into resilience measures and objective functions using the proposed transformation mechanism. The proposed transformation and modeling approaches aim to address the challenging task of explicitly integrating the notion of "resilience" into quantitative and systematic decision making.

To the best of our knowledge, our work is the first to quantify resilience of decentralized infrastructure to risks due to sea level rise and introduces a model-based approach that could be generalized to any other infrastructure system subject to various environmental risks. An important insight that we obtain from our approach indicates that a system that is able to resist disruptions is not necessarily resilient when it comes to addressing impact propagation to other critical infrastructure systems. We implemented our methodology using a case study based on Miami-Dade County's OSTDS system. Our analysis using the proposed approach reveals that nearly 8% of the septic sites need immediate attention for adaptation in this region. Future extensions to this work include the application of this methodology for assessing resilience of other standalone infrastructure systems such as coastal defense systems, agricultural fields and historical sites.